\documentclass[a4paper,12pt]{article}

\usepackage[utf8]{inputenc}
\usepackage[T1]{fontenc}
\usepackage[margin=2.5cm]{geometry}
\usepackage{amsmath,amssymb}
\usepackage{mathtools}
\usepackage{graphicx}
\usepackage{setspace}
\usepackage{microtype}
\usepackage{float}
\usepackage{caption}

\usepackage{hyperref}
\usepackage{doi}

\usepackage[
    backend=biber,
    doi=true,
    url=true,
    hyperref=true
]{biblatex}

\title{%
    Partition Space Maps for Community Detection: Visualizing Algorithmic Behaviour and Guiding Search
}
\author{
    Fabio~Morea$^{1,*}$\\[4pt]
    {\small $^{1}$Area Science Park, Trieste, Italy}\\
    {\small $^{*}$Corresponding author: \texttt{fabio.morea@areasciencepark.it}}
}

\date{}

\begin{document}
\maketitle

\begin{abstract}  
Community-detection methods search over possible partitions of a network, but the structure of this partition space is rarely examined directly. This paper introduces two complementary tools that make~$\mathcal{P}$ analytically and visually accessible.
First, a canonical labelling scheme based on the Restricted Growth Sequence~(RGS) is adopted, assigning each partition a unique, permutation-invariant identifier and eliminating the need for pairwise similarity measures such as the Normalized Mutual Information.
Second, granularity~$\Gamma$ is defined and combined with modularity~$Q$ to produce a two-dimensional projection: the Partition Space Map~(PSM).
Within the $(\Gamma,Q)$ plane, only a discrete set of integer values of~$\Gamma$ is attainable, and upper and lower bounds on~$Q$ are derived directly from the deviation matrix.
These tools are demonstrated through complete enumeration of~$\mathcal{P}$ for small benchmark graphs, revealing the fine structure of modularity degeneracy and the combinatorial constraints that shape the feasible region.
The framework provides an exact, assumption-free reference for evaluating heuristic community detection algorithms and lays the groundwork for geometry-aware exploration of partition space in larger networks.
\end{abstract}

\section{Introduction}
Identifying meaningful partitions of a network—typically interpreted as communities, i.e., groups of nodes more densely connected internally than externally—remains an active area of research in network science~\cite{Newman2004, fortunato2016userguide, hric2014comparative}.
The field is characterised by a diversity of algorithms and modelling paradigms. Recent reviews have documented this methodological landscape~\cite{fortunato2016userguide} and observed that algorithmic outputs can vary substantially even on controlled benchmarks~\cite{hric2014comparative,lancichinetti2012consensus}.  

Most community detection methods are framed as optimisation procedures that seek partitions maximising a quality function, commonly the Newman–Girvan modularity~$Q$~\cite{Newman2006}.  
Although modularity optimisation has proven useful across domains, certain fundamental properties of the partition space itself present challenges for interpretation.  
The set of all partitions, denoted~$\mathcal{P}$, grows with the Bell number~$B_n$ and is combinatorial, discrete, and non-Euclidean.  
Searching~$\mathcal{P}$ therefore differs from classical optimisation in continuous spaces: the landscape is large, discrete, and not easily characterised by local heuristics alone.

In practice, most algorithms return a single high-modularity partition but provide limited information about the broader structure of the solution space.  
Questions such as how many near-optimal partitions exist, how they are distributed, and to what extent the modularity optimum exhibits degeneracy often remain unexplored.  
The resolution limit of modularity~\cite{Fortunato2007, Lancichinetti2011, Darst2014}, together with evidence for exponentially many near-degenerate high-modularity partitions~\cite{Good2010}, suggests that focusing on a single solution may overlook important structural information, particularly for networks beyond minimal size.

 These observations point to potential value in a structured representation of~$\mathcal{P}$ that could help researchers understand how algorithms explore the solution landscape, assess the influence of initialisation choices, and identify when multiple structurally distinct partitions achieve comparable quality scores. 

This work introduces two complementary tools to explore the partition space.
First, we adopt the Restricted Growth Sequence~(RGS) as a canonical, permutation-invariant labelling of partitions, enabling exact comparisons without pairwise similarity measures.
Second, we define granularity~$\Gamma$ and use it together with modularity~$Q$ to construct the Partition Space Map~(PSM), a two-dimensional projection of~$\mathcal{P}$ onto the $(\Gamma, Q)$ plane.
The PSM includes closed-form upper and lower envelopes for modularity, derived directly from the network's deviation matrix and independent of specific partition structures.
Together, these tools provide a diagnostic framework---not a new detection method---that makes the global organisation of~$\mathcal{P}$ visually and analytically accessible, and offers computable reference bounds against which the output of any heuristic algorithm can be evaluated.

\section{Methods}
\subsection{Graphs, Communities, and Partitions}

Consider an undirected graph $G=(V, E)$ comprising a set of $n$~vertices $V$
and an edge set~$E$. While the vertex set possesses no inherent ordering,
any transition to a matrix-based formalism necessitates the adoption of a
fixed indexing scheme, $V = \{v_1, \dots, v_n\}$. This choice is treated
here as a strictly conventional coordinate system---a practical requirement
for computation rather than a structural assumption.

Under this convention, the symmetric weight matrix $\mathbf{W} = [w_{ij}]$
is defined via a non-negative weight function $w\colon E \to \mathbb{R}_{>0}$,
where $w_{ij} = w(v_i, v_j)$ if $\{v_i, v_j\} \in E$, and zero otherwise.
The strength of node~$i$, denoted $s_i = \sum_{j} w_{ij}$, provides a
measure of its local connectivity; in the unweighted case, unit weights are
assumed, such that $s_i$ reduces to the standard node degree~$d_i$.

One may impose a variety of structures onto the graph by defining a
\emph{partition} $\Pi = \{C_1, \dots, C_K\}$ of the vertex set. To be
formally well-posed, any such partition must satisfy the requirements of
disjointness ($C_i \cap C_j = \varnothing$ for $i \neq j$), covering
($\bigcup_k C_k = V$), and non-emptiness ($C_k \neq \varnothing$ for
all~$k$). In computational practice, a partition is represented through a
membership vector $\mathbf{c} = [c_1, \dots, c_n]$, where each entry~$c_i$
denotes the label assigned to node~$i$.

To distinguish meaningful clusters from the vast space of arbitrary
partitions, let $W_{\mathrm{in}}(C)$ denote the total weight of edges
internal to a subset~$C$ and $W_{\mathrm{out}}(C)$ the weight of edges
connecting~$C$ to the remainder of the graph. Following~\cite{radicchi2004},
a subset~$C_k$ is identified as a \emph{community} if its internal
connectivity exceeds its external connectivity:
$W_{\mathrm{in}}(C_k) > W_{\mathrm{out}}(C_k)$.

\begin{figure}[h]
    \centering
    \includegraphics[width=\linewidth]{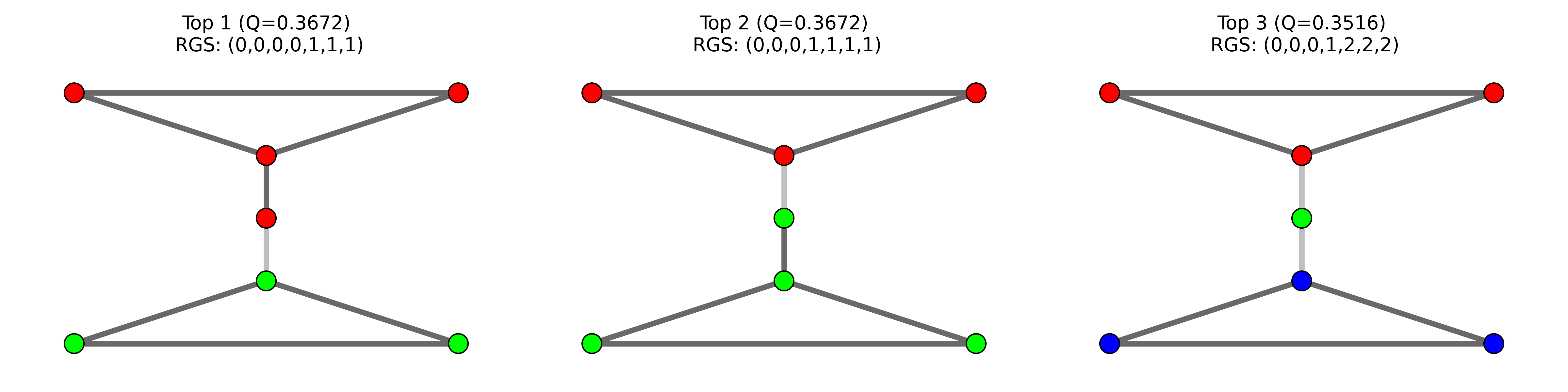}
    \caption{%
        A test network consisting of two cliques connected by a single bridge node.
        This minimal configuration contains only seven nodes, which allows the enumeration of all possible partitions by brute force and thus obtain exact ground-truth modularity values.
    }
    \label{fig:test-network}
\end{figure}

The benchmark network illustrated in Figure~\ref{fig:test-network} serves
as a primary example throughout this study. Though modest in
scale---permitting exhaustive brute-force enumeration of all possible
partitions---the network is sufficiently complex to exhibit two distinct
communities and a characteristically ambiguous node.
\subsection{Modularity}

Modularity assigns a quality score to a partition by measuring how much the density of edges within communities exceeds the expectation under a null model.
In the standard Newman--Girvan formulation~\cite{Newman2006}, modularity is defined as
\begin{equation}
    Q(\Pi)
    = \frac{1}{2m}
      \sum_{i,j}
      \left(
        w_{ij}
        - \frac{s_i\, s_j}{2m}
      \right)
      \delta(c_i,c_j),
    \label{eq:modularity}
\end{equation}
where $\delta(c_i,c_j)=1$ if $c_i=c_j$ and $0$ otherwise, $s_i=\sum_j w_{ij}$ is the strength of node~$i$, and $2m=\sum_{i,j} w_{ij}$ is the total edge weight of the network. 
In matrix form, modularity can be written as
\[
    Q = \frac{1}{2m} \sum_{i,j} X_{ij}\,\Delta_{ij},
\]
with
\begin{equation}
    N = \frac{s\,s^\top}{2m}, \qquad
    X = W - N, \qquad
    \Delta_{ij} = \delta(c_i,c_j).
    \label{eq:matrices}
\end{equation}
The deviation matrix~$X$ is a fixed property of the graph: each entry $X_{ij}$ quantifies whether the observed weight between nodes $i$ and~$j$ exceeds ($X_{ij}>0$) or falls short of ($X_{ij}<0$) the expectation imposed by the null model.

\begin{figure}[h]
    \centering
    \includegraphics[width=\linewidth]{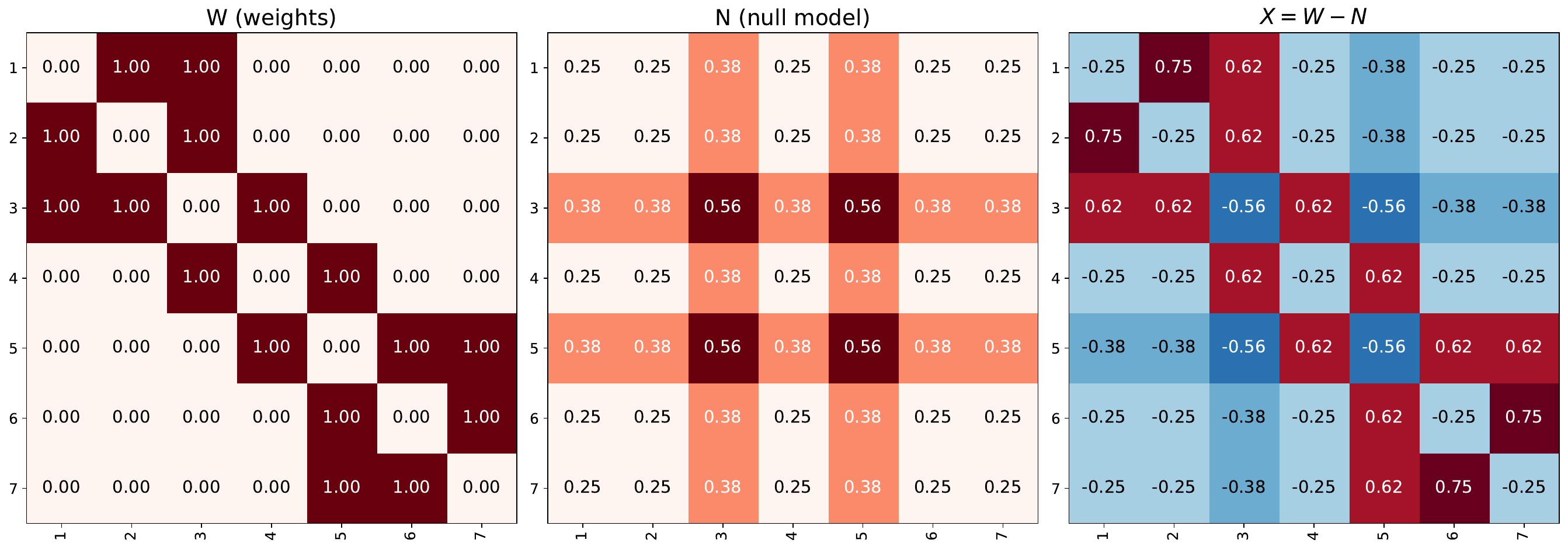}
    \caption{%
        Matrices $W$, $N$, and $X$ for the test network, an unweighted graph with $n=7$ nodes and $m=8$ edges.
        Left: weight matrix~$W$ (red cells indicate the presence of edges).
        Centre: null-model matrix $N = s\,s^\top / 2m$.
        Right: deviation matrix $X = W - N$.
    }
    \label{fig:matrices-example}
\end{figure}

The indicator matrix~$\Delta$ acts as a partition-dependent mask, restricting the sum to block-diagonal entries corresponding to intra-community node pairs.
Modularity can therefore be interpreted as the normalized sum of within-community deviations from the null-model expectation.
To illustrate the approach, we apply it to the test network shown in Figure~\ref{fig:test-network}, and the corresponding matrices are reported in Figure~\ref{fig:matrices-example}.

\begin{figure}[h]
    \centering
    \includegraphics[width=\linewidth]{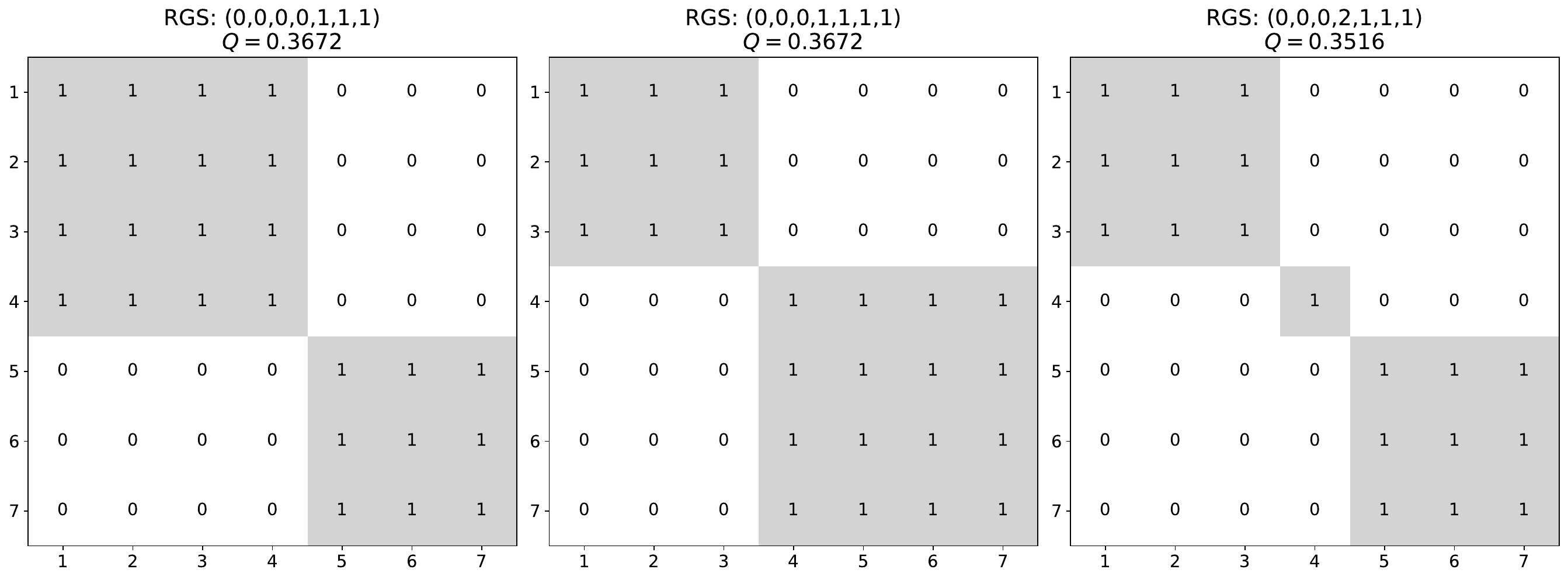}
    \caption{%
        Indicator matrix~$\Delta$ for three partitions of the example graph, with RGS encodings and $(\Gamma,Q)$ coordinates.
        Left and centre: two structurally distinct partitions sharing $\Gamma=25$ and $Q=0.3672$, illustrating modularity degeneracy.
        Right: a three-community partition with $\Gamma=19$ and $Q=0.3516$.
        Grey cells mark entries with $\Delta_{ij}=1$; graph insets show the community assignment.
    }
    \label{fig:delta}
\end{figure}

Modularity is bounded above and below by values determined solely by the sorted entries of $X$. Every partition places all $n$ diagonal entries $X_{ii}$ inside $\Delta$ (each node shares a community with itself), giving a fixed diagonal contribution
\begin{equation}
    Q_{\mathrm{diag}} = \frac{1}{2m} \sum_{i=1}^{n} X_{ii}.
    \label{eq:Qdiag}
\end{equation}
For graphs without self-loops, $X_{ii}=-s_i^2/(2m)$, so $Q_{\mathrm{diag}} < 0$.

The remaining off-diagonal entries in $\Delta$ are drawn from the upper triangle of $X$, with each symmetric pair $(i,j)$ contributing $2X_{ij}$. 
To construct the tightest upper bound, sort the off-diagonal entries in descending order and form the cumulative sum
\begin{equation}
    Q^{\max}(\ell) 
    = \frac{1}{2m} \Biggl( \sum_i X_{ii} + 2 \sum_{t=1}^{\ell} X_{\sigma(t)} \Biggr),
    \label{eq:upper-cumulative}
\end{equation}
where $\sigma$ is the permutation sorting the off-diagonal entries and $\ell = 0,1,\ldots,\binom{n}{2}$ counts how many pairs have been included. 
The corresponding granularity is $\Gamma = n + 2\ell$, and the curve $(\Gamma(\ell), Q^{\max}(\ell))$ traces the tightest achievable upper bound.

The lower bound is constructed analogously by sorting in ascending order:
\begin{equation}
    Q^{\min}(\ell) 
    = \frac{1}{2m} \Biggl( \sum_i X_{ii} + 2 \sum_{t=1}^{\ell} X_{\sigma'(t)} \Biggr),
    \label{eq:lower-cumulative}
\end{equation}
where $\sigma'$ sorts the off-diagonal entries ascendingly. 
Together, the two cumulative curves define a closed region in the $(\Gamma,Q)$ plane that contains all realisable partitions.

 \subsection{Granularity}

 To characterize the structure of each partition $\Pi \in \mathcal{P}$, we introduce the notion of \emph{granularity}.
Using the partition matrix $\Delta$ defined above, whose entries indicate whether two nodes belong to the same community, we define the granularity of a partition as the total number of ones in this matrix:
\begin{equation}
\Gamma(\Pi) = \sum_{i=1}^{n}\sum_{j=1}^{n} \Delta_{ij}.
\label{eq:granularity}
\end{equation}

This quantity measures the total mass of intra-community relations encoded by the partition.
Because $\Delta_{ij}=1$ whenever nodes $i$ and $j$ belong to the same community, the above definition admits an equivalent expression in terms of community sizes.
If the partition consists of communities $C_1,\ldots,C_K$ with sizes $|C_k|$, then each community contributes $|C_k|^2$ ones to $\Delta$, yielding
\begin{equation}
\Gamma(\Pi) = \sum_{k=1}^{K} |C_k|^2 .
\label{eq:level}
\end{equation}

Granularity is bounded by two topological extremes: the all-singleton partition, where $\Gamma = n$, and the all-in-one partition, where $\Gamma = n^2$.
Importantly, the interval $[n,n^2]$ is not continuously populated. Rather, only those values that can be written as a sum of squares corresponding to an integer composition of $n$ are attainable.
This discreteness is therefore not a computational artifact but an intrinsic structural property of the partition space $\mathcal{P}$.

\begin{figure}[h]
    \centering
    \includegraphics[width=\linewidth]{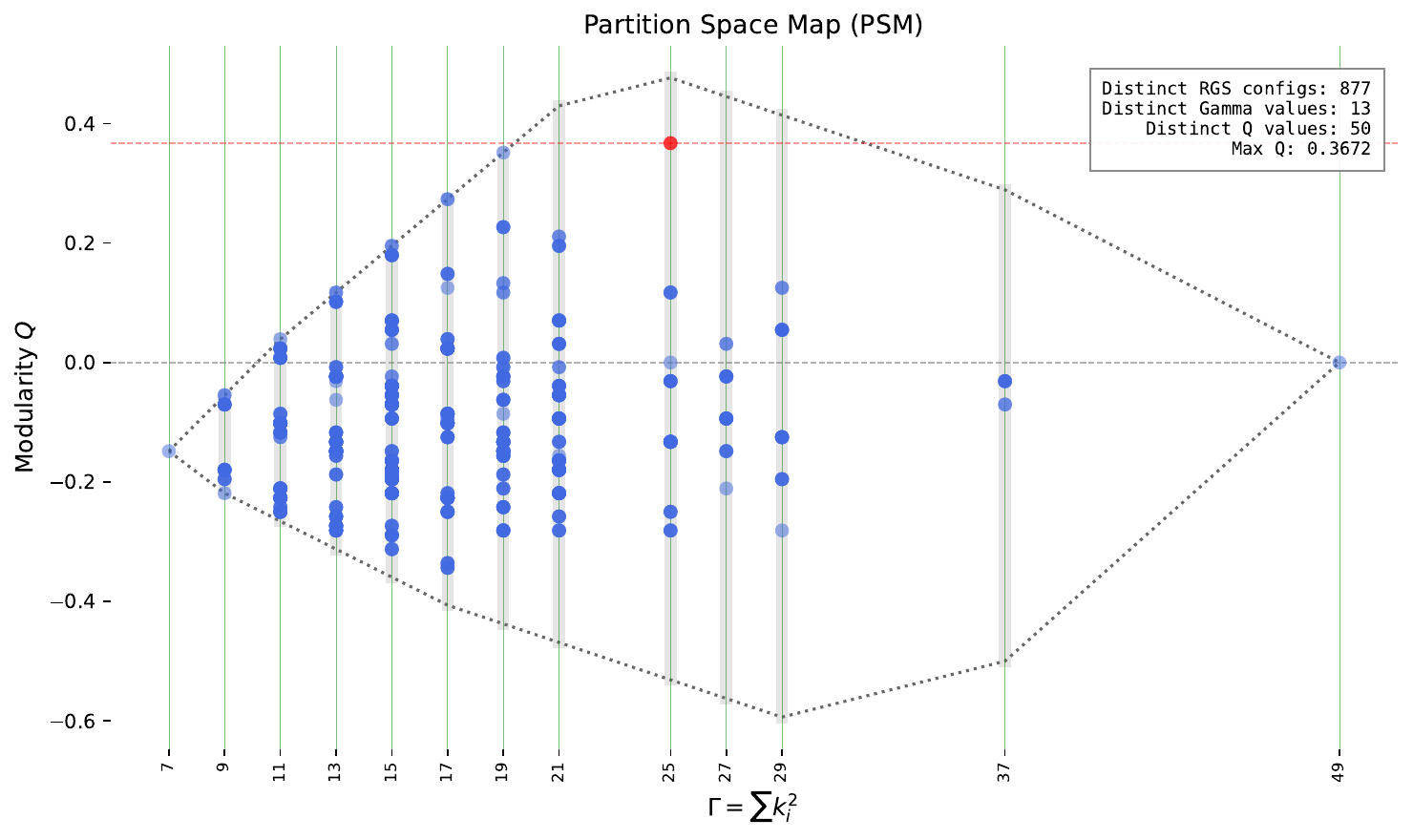}
    \caption{%
        Partition Space Map of the example graph ($n=7$).
        Each point is a canonical partition projected onto~$(\Gamma,Q)$.
        The red dot marks maximum modularity; dashed lines indicate the upper and lower modularity bounds.
        Only 13~distinct $\Gamma$-values and 50~distinct $Q$-values exist among all 877~partitions, with many points overlapping.
    }
    \label{fig:psm}
\end{figure}

By projecting every feasible partition onto the $(\Gamma, Q)$ plane, one constructs a two-dimensional atlas of $\mathcal{P}$—hereafter referred to as the \emph{Partition Space Map} (PSM). The PSM of the benchmark network, illustrated in Fig.~\ref{fig:psm}, reveals the organisational landscape of the graph, mapping the interplay between community quality and structural resolution.

The left anchor ($\Gamma=n$, i.e. each node is assigned to a different community) has modularity $Q_{\mathrm{sing}}=(2m)^{-1}\sum_i X_{ii}$, which is strictly negative for graphs without self-loops.
The right anchor ($\Gamma=n^2$, all nodes assigned to the same partition) satisfies $Q=0$ identically, since $\sum_{i,j}X_{ij}=0$; it is therefore the only topology-invariant reference point in any PSM.

\clearpage
\subsection{Canonical labelling via Restricted Growth Sequences}
Canonical representations of set partitions are well known in combinatorics, but their use as permutation-invariant labels for community partitions has received little attention in network science. 
Because community labels are arbitrary, different label vectors can encode the same partition; for example, $(0,0,1,1)$ and $(1,1,0,0)$ both describe identical two-community structures. 
Comparing partitions therefore typically requires either solving a label-matching problem or computing a permutation-invariant similarity measure such as the Normalized Mutual Information~(NMI), which can be computationally costly.

\begin{figure}[h]
    \centering
    \includegraphics[width=\linewidth]{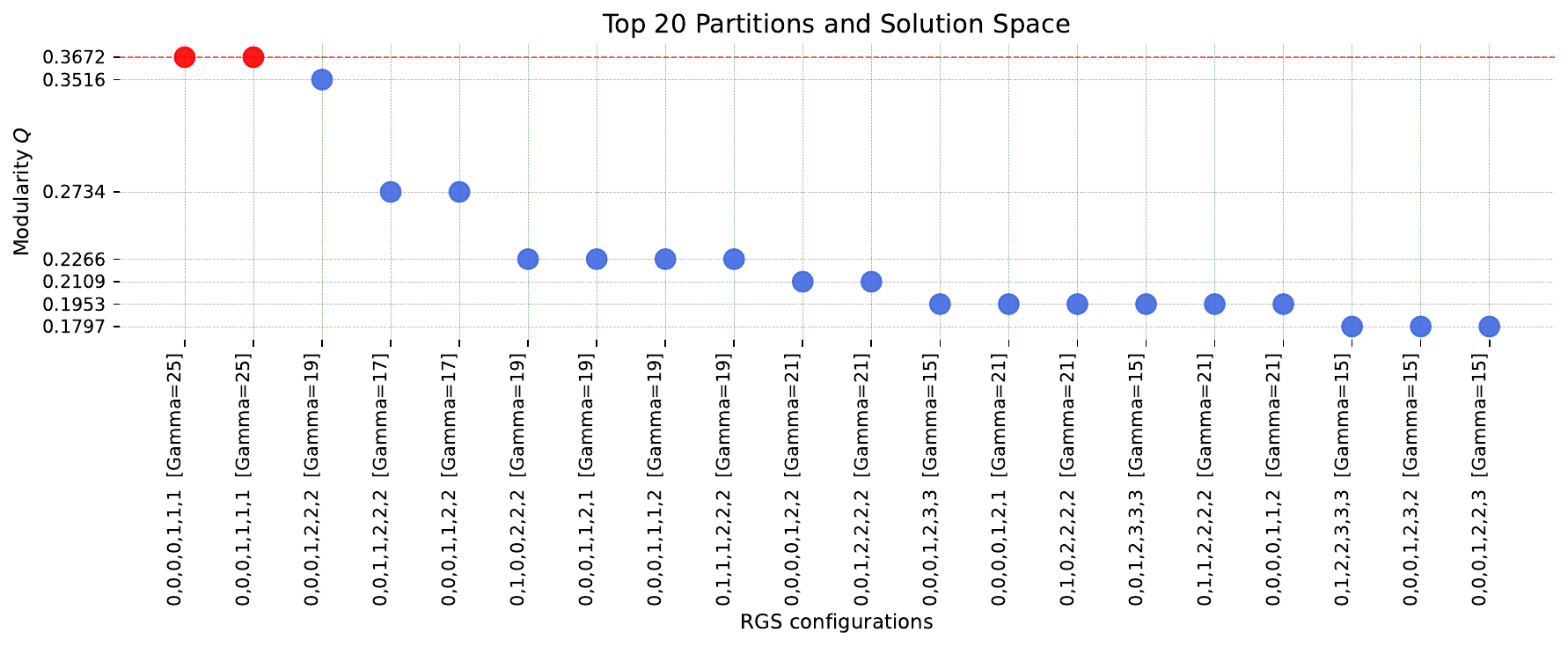}
    \caption{%
       Top 20 partitions of the example graph visualized in the Partition Space Map. Each point represents a distinct partition, with its modularity $Q$ on the vertical axis and the canonical RGS configuration on the horizontal axis. Red points indicate globally optimal partitions. Ordering via RGS provides a permutation-invariant representation that clarifies the existence of modularity plateaus.
       } 
    \label{fig:5}
\end{figure}

We resolve this by adopting the Restricted Growth Sequence~(RGS), a classical combinatorial encoding\cite{Nijenhuis1978} that provides each partition with a unique canonical form. The RGS is constructed by scanning the membership vector from left to right and re-indexing labels to $\{0,1,2,\ldots\}$ in order of first appearance.
Formally, given membership $(c_1,\ldots,c_n)$, the RGS $(r_1,\ldots,r_n)$ satisfies $r_1=0$ and, for $i>1$, $r_i=r_j$ if $c_i=c_j$ for some $j<i$, or $r_i=1+\max\{r_1,\ldots,r_{i-1}\}$ otherwise.
The encoding is bijective---every distinct partition maps to exactly one RGS---and satisfies the restricted-growth constraint $r_i\leq 1+\max\{r_1,\ldots,r_{i-1}\}$.
Two partitions are identical if and only if their RGS vectors coincide, reducing the typical $O(n \log n)$ cost of label-matching or permutation-invariant comparisons to $O(n)$. Figure~\ref{fig:5} illustrates the use of the RGS encoding to organize distinct partitions and display their corresponding modularity values.

\clearpage
\section{Examples of applications}

\subsection{Exact Characterisation of a Minimal Benchmark}

The utility of the PSM is demonstrated using the unweighted graph illustrated in Fig.~\ref{fig:matrices-example} ($n=7$, $m=8$). The associated partition space $\mathcal{P}$ comprises 877 distinct partitions, each uniquely identified by its RGS. To visualize the microscopic structure of these partitions, Fig.~\ref{fig:delta} presents the indicator matrices for three selected candidates.

The first two partitions---$\mathbf{c}_1 = (0,0,0,0,1,1,1)$ and $\mathbf{c}_2 = (0,0,0,1,1,1,1)$---share a granularity of $\Gamma=25$ and achieve a global modularity maximum of $Q=0.3672$. This instance exemplifies the phenomenon of algorithmic degeneracy: structurally distinct partitions may yield identical quality scores, complicating the identification of a unique ``ground truth'' community structure. A third candidate, $\mathbf{c}_3 = (0,0,0,2,1,1,1)$, introduces an additional community, shifting the resolution to $\Gamma=19$ with a corresponding modularity of $Q=0.3516$. When viewed globally, the 877 partitions project onto a sparse set of only 13 distinct values of $\Gamma$ and 50 distinct values of $Q$. This compression confirms a pervasive degeneracy across both coordinates. The density of partitions peaks at intermediate values of $\Gamma$ reflects the underlying distribution of the partition space, where configurations consisting of balanced community sizes are statistically most numerous. Conversely, the all-singleton partition at $\Gamma=7$ exhibits negative modularity, a result consistent with the absence of self-loops in the underlying graph. Crucially, the modularity envelopes (marked by dotted lines) explicitly demarcate the feasible region of the partition space.

This visualisation confirms that the search for high-modularity partitions is bounded not merely by algorithmic heuristics, but by hard topological limits inherent to the deviation matrix.

\clearpage
\subsection{PSM as a tool to visualise algorithm behaviour}\label{sec:lfr}

PSM is a useful tool to evaluate how heuristic community detection algorithms navigate the partition space. In this example it is applied to a standard LFR benchmark graph \cite{lancichinetti2008benchmark} with $n = 50$ nodes, a mixing parameter of $\mu = 0.25$ and random weights. This construction provides a known planted community structure, which serves as a fixed reference point within the $(\Gamma, Q)$ plane.

\begin{figure}[h]
  \centering
  \includegraphics[width=\textwidth]{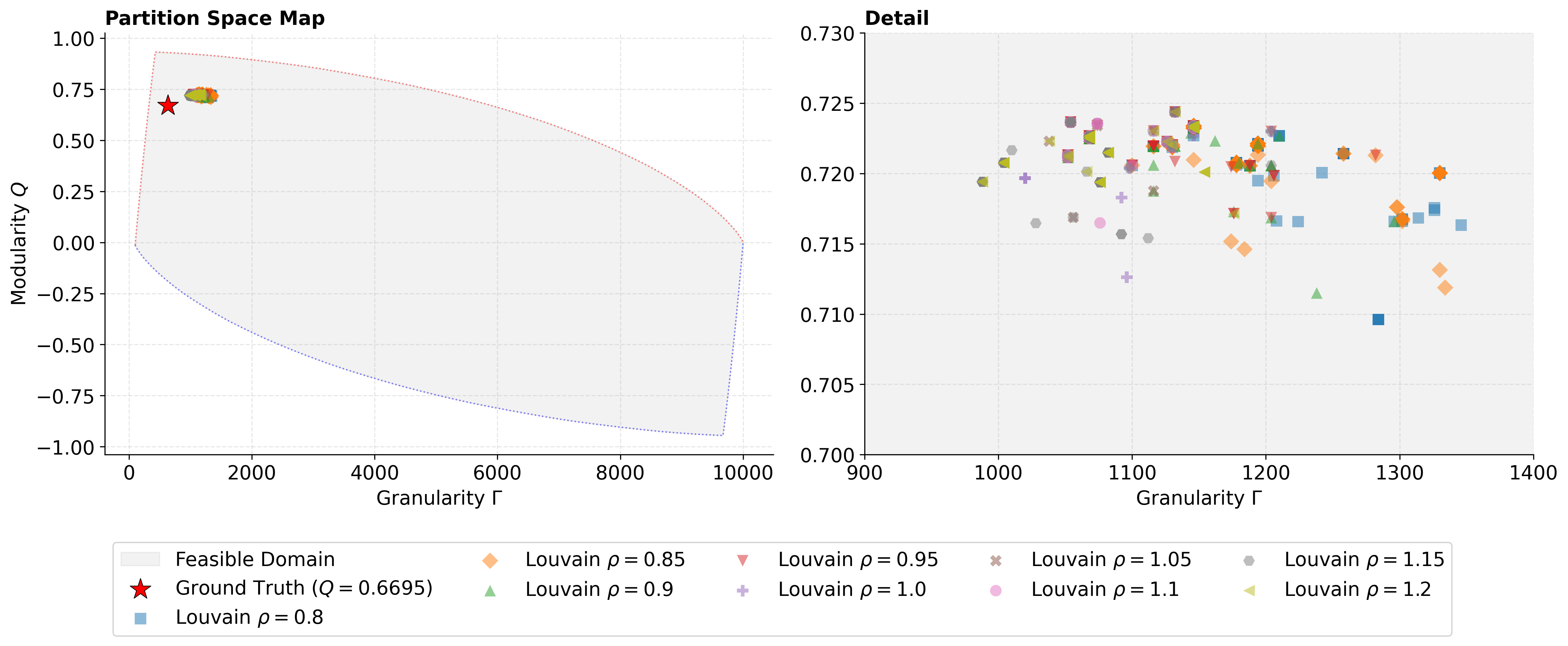}
  \caption{Partition Space Map for a weighted LFR graph ($n=50$, $\mu=0.25$) showing how the Louvain algorithm's output shifts across the feasible region as the resolution parameter varies. The ($\Gamma$, $Q$) projection makes the trade-off between granularity and modularity geometrically explicit: each resolution value produces a distinct cluster of unique partitions and their spread reveals both the direction and magnitude of the algorithm's sensitivity to parameter value.}
  \label{fig:louvain_resolutions}
\end{figure}

\paragraph{The Response of Louvain to Resolution Variation.}
Figure~\ref{fig:louvain_resolutions} illustrates the projection of partitions generated by the Louvain algorithm across a range of nine resolution parameters, $\rho \in \{0.80, 0.85, \dots, 1.20\}$. For each value of $\rho$, 100 independent runs were conducted with randomized node orderings; distinct partitions were subsequently identified and retained through their unique RGS encoding.

As expected, all Louvain outputs are contained strictly within the feasible region defined by the modularity envelopes. The distribution of these points follows a systematic trajectory as a function of $\rho$: higher resolution values shift the resulting partition cloud toward smaller $\Gamma$ (favouring a larger number of smaller communities), whereas lower values bias the search toward coarser structures with higher $\Gamma$. 

Notably, the planted ground truth occupies a central position relative to this family of solutions. This mapping reveals the sensitivity of the optimisation process: while certain resolution values successfully approximate the ground-truth neighbourhood, others diverge significantly in both modularity and structural granularity, illustrating the inherent challenge of parameter selection in heuristic search.

\begin{figure}[h]
  \centering
  \includegraphics[width=\textwidth]{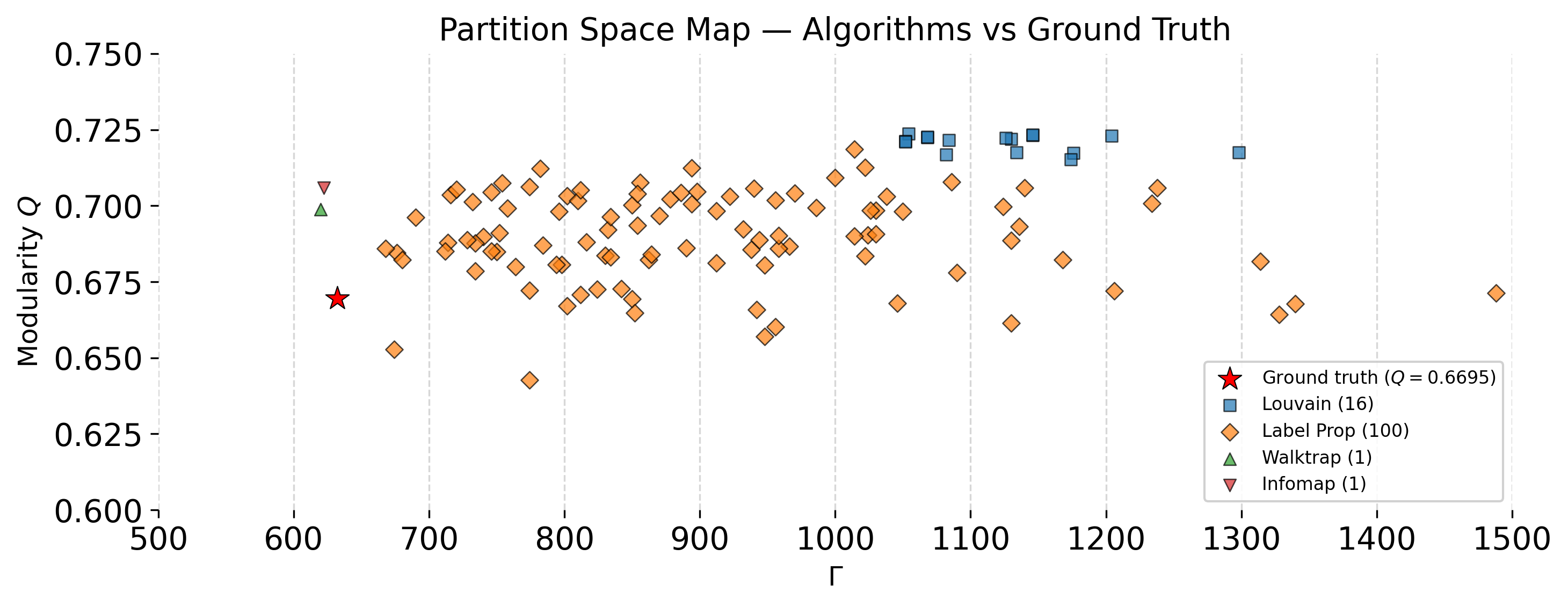}
  \caption{Partition Space Map comparing the results of 100 independent runs of four algorithms on the same weighted LFR graph. The ($\Gamma$, Q) projection reveals not only how much algorithmic outputs differ in quality, but in which direction—toward finer or coarser granularity—they diverge from the ground truth (star) and from each other, providing a geometric characterisation of algorithmic behaviour that scalar modularity scores alone cannot capture.}
  \label{fig:algorithm_comparison}
\end{figure}

\paragraph{Comparison of four algorithms.}
Figure~\ref{fig:algorithm_comparison} compares Louvain ($\rho=1$), Label Propagation, Infomap, and Walktrap, each run 100 times with random node
permutations. Louvain and Infomap show compact clusters close to the ground truth, indicating
limited sensitivity to initialisation.  
Label Propagation spans a noticeably wider region in both $\Gamma$ and $Q$,
reflecting its well-known instability.  
Walktrap, being effectively deterministic, produces very few distinct
partitions that occupy a narrow band of the feasible region.
Overall, the PSM reveals qualitative differences that are not visible when
comparing only modularity scores: algorithms may return partitions with similar
$Q$ but substantially different structural organisation as captured by $\Gamma$.

\clearpage

\paragraph{Visualisation of algorithmic trajectories.}

The Partition Space Map provides a natural framework for visualizing the trajectories of community detection algorithms. Algorithms that generate a sequence of partitions through progressive aggregation or splitting of communities naturally trace a trajectory in partition space. 

To illustrate this idea, we consider the Louvain algorithm and record the sequence of partitions produced during its greedy optimisation of modularity. Each run starts from the singleton partition, where every node forms its own community, corresponding to the minimum attainable granularity $\Gamma = n$. As nodes are progressively aggregated into communities, $\Gamma$ increases while modularity $Q$ typically improves, producing a trajectory that moves toward regions of higher modularity and larger community sizes.

\begin{figure}[h]
\centering
\includegraphics[width= \textwidth ]{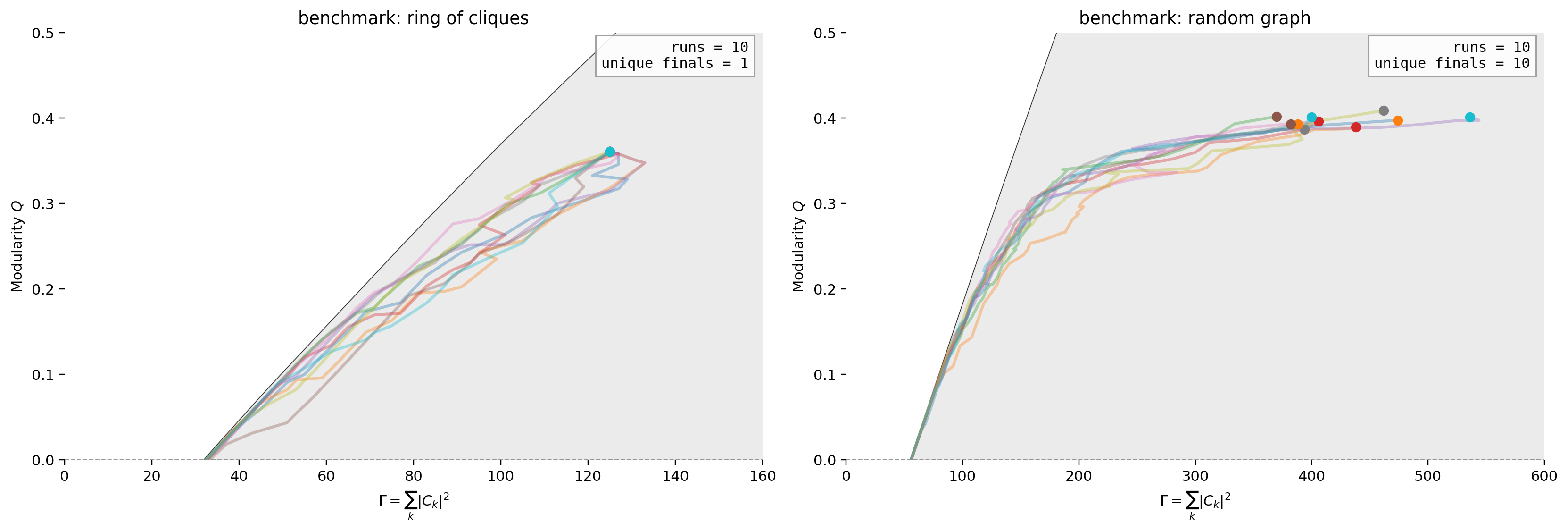}
\caption{
Trajectories of the Louvain algorithm in the Partition Space Map. Each  line represents the sequence of partitions generated during a single run of the algorithm, starting from the singleton partition. 
(\textbf{Left}) Ring of cliques benchmark with a clear community structure. 
(\textbf{Right}) Random graph without a clear modular structure. 
Here the trajectories diverge and terminate at distinct local optima, 
revealing multiple competing solutions with different granularities. 
}
\label{fig:louvain_trajectories}
\end{figure}

Figure~\ref{fig:louvain_trajectories} shows two examples. 
In the left panel, we consider a \emph{ring of cliques}, a benchmark network with a clear and essentially unique community structure. Although the intermediate trajectories vary slightly across runs as a result of the greedy nature of the algorithm, all runs converge to the same final partition.

In contrast, the right panel shows the behaviour of Louvain on a random graph without a clear community structure. Here, the trajectories diverge and terminate at distinct local optima with different granularities, revealing multiple competing solutions.

These examples illustrate how the PSM can be used to visualize both the dynamics and the stability of community detection algorithms.

\clearpage
\subsection{How partition space varies with network density}

To demonstrate the fundamental relationship between network structure and partition space geometry, we construct a controlled sequence of benchmark networks that interpolate between two extremes: a ring of cliques (minimal density, sharp community structure) and a complete graph (maximum density, no community structure). We begin with $n=50$ vertices divided into $k=5$ equal-sized cliques, where each clique is internally complete and consecutive cliques are connected by a single bridge edge forming a ring topology. An interpolation parameter $\alpha \in [0,1]$ controls the addition of inter-community edges: at $\alpha=0$, only the ring structure exists; at $\alpha=1$, all possible edges are present, yielding a complete graph. For intermediate values of $\alpha$, we add a fraction $\alpha$ of all possible inter-community edges, selected uniformly at random. This construction guarantees monotonic convergence from structured communities to complete homogeneity while maintaining a well-defined ground truth partition throughout.

\begin{figure}[h]
\centering
\includegraphics[width=\textwidth]{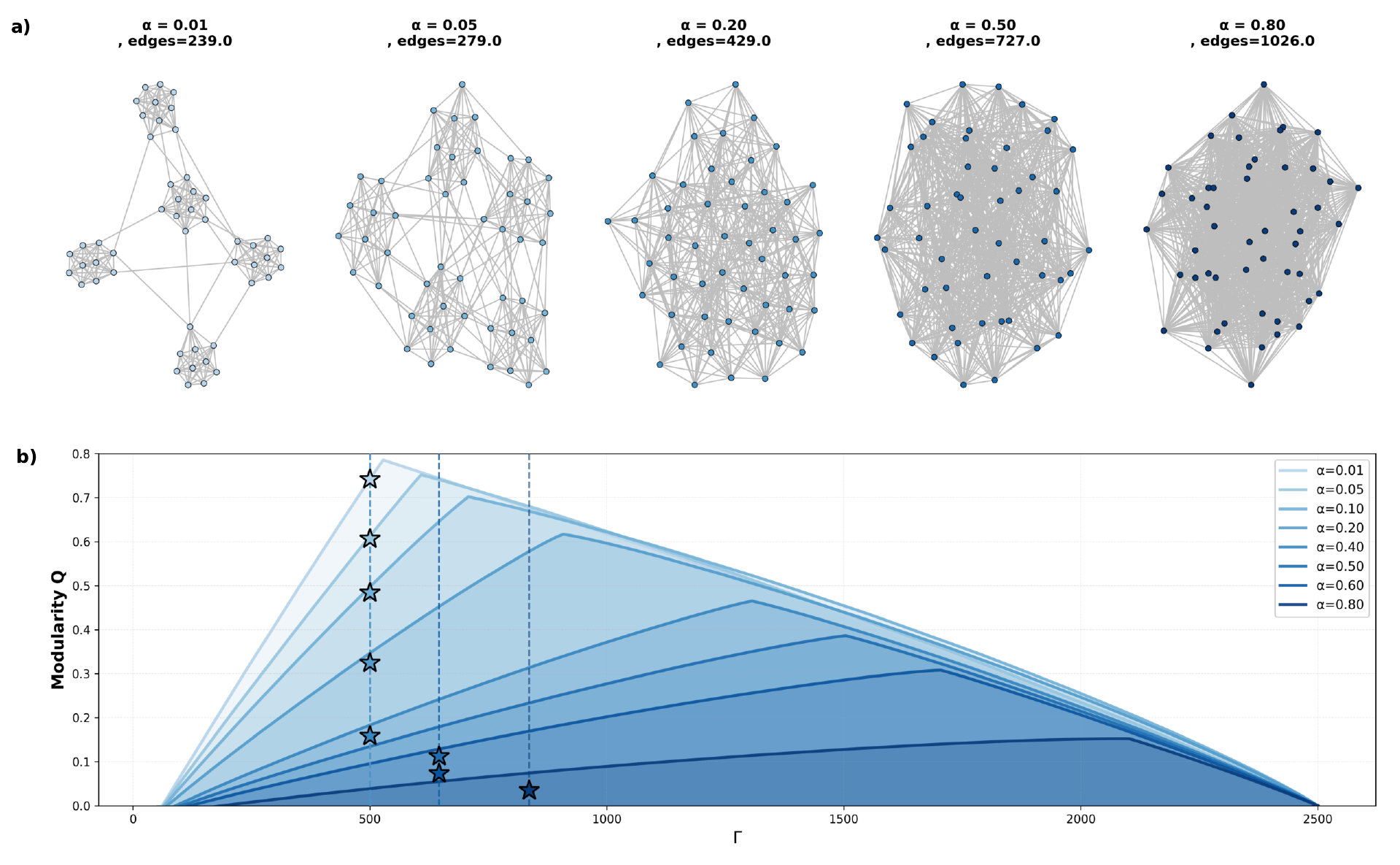}
\caption{Partition space collapse as network structure changes. (a)~Network evolution from ring of cliques ($\alpha=0.01$, light blue) to complete graph ($\alpha=0.80$, dark blue). (b)~Upper modularity bounds projected onto the $(\Gamma,Q)$ plane for networks with varying~$\alpha$. Stars mark best Louvain solutions (50 trials); vertical dashed lines indicate granularity of these solutions. The size of partition space systematically decreases as community structure degrades.}
\label{fig:partition_collapse}
\end{figure}

Figure~\ref{fig:partition_collapse} presents both the structural evolution of these networks (panel a) and the corresponding collapse of their partition spaces (panel b). The network visualisations clearly show the progressive increase in density from sparse ring structure ($\alpha=0.01$) to near-complete connectivity ($\alpha=0.80$). Panel b displays the upper modularity envelopes for eight networks that span $\alpha \in [0,1]$. The shaded region under each envelope represents the feasible partition space. As $\alpha$ increases, this region systematically collapses. Star markers indicate the best modularity achieved by the Louvain algorithm in 50 independent runs, with vertical dashed lines marking the corresponding granularity $\Gamma$. Initially, algorithmic solutions are aligned to $\Gamma = 500$ (i.e. 5 communities) but as density increases, and structure weakens, the solution collapses into larger communities, illustrating both the degradation of community structure and the increasing difficulty of the detection problem.

\clearpage
\section{Area Between the Curves}

The geometry of the Partition Space Map (PSM) can be summarized by a scalar quantity that measures the size of the feasible region. To obtain a scale-independent measure, we introduce the normalized granularity coordinate
\[
\tilde{\Gamma} = \frac{\Gamma - n}{n^2 - n},
\]
which maps the attainable range of $\Gamma$ to the interval $\tilde{\Gamma}\in[0,1]$. Expressing the modularity bounds as functions of this normalized coordinate, we define the \emph{area between the curves} as

\[
ABC =
\int_{0}^{1}
\left[
Q_{\mathrm{upper}}(\tilde{\Gamma}) -
Q_{\mathrm{lower}}(\tilde{\Gamma})
\right] d\tilde{\Gamma}.
\]

Because both axes are dimensionless in the normalized coordinates, $ABC$ is itself dimensionless and directly comparable across networks of different sizes.
Small values of $ABC$ correspond to narrow feasible regions, indicating that the modularity of possible partitions is strongly constrained across the granularity axis. Larger values correspond to broader feasible regions, reflecting a wider range of attainable modularity values, and therefore a richer set of competing partitions.

From this perspective, $ABC$ can be interpreted as a coarse measure of the structural complexity of the partition space. Networks with small $ABC$ exhibit tightly constrained partition landscapes, whereas networks with large $ABC$ display broader and potentially more complex partition spaces.

\begin{figure}[h]
    \centering
    \includegraphics[width=\linewidth]{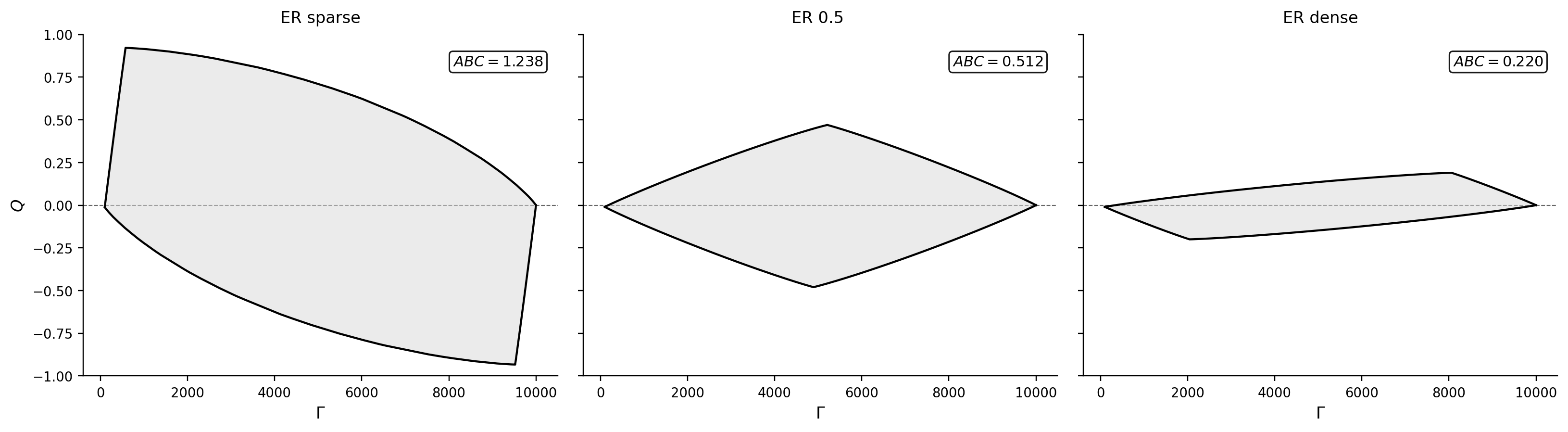}
    \caption{
    Illustration of the \emph{area between the curves} ($ABC$) in the Partition Space Map. 
    The shaded region represents the feasible set of partitions in the $(\tilde{\Gamma},Q)$ plane, bounded by the upper and lower modularity envelopes. 
    The quantity $ABC$ is defined as the normalized area between these curves across the full granularity range. 
    This scalar summarizes the overall width of the feasible region and provides a compact measure of how strongly the modularity of partitions is constrained by the network structure.
    }
    \label{fig:abc_psm}
\end{figure}

Figure~\ref{fig:abc_psm} illustrates the construction of this measure. The shaded region represents the feasible portion of the PSM bounded by the upper and lower modularity envelopes. The value of $ABC$ corresponds to the total area enclosed between these curves once the granularity axis is normalized. In practice, this quantity can be estimated numerically from sampled partitions or from empirical approximations of the bounding envelopes.

\clearpage
\section{PSM-inspired two-phase modularity maximisation}
\label{sec:psm_two_phase}

The Partition Space Map is introduced in this work primarily as a descriptive representation of partition space. However, the same geometric construction also suggests a practical search strategy for modularity maximisation.  The central idea is that repeated heuristic runs often reveal \emph{stable co-assignment relations} among subsets of nodes. When such relations are consistently detected  across high-quality partitions, they can be treated as reliable structural information and used to restrict the subsequent search to a smaller, more relevant portion of the feasible region in the $(\Gamma,Q)$ plane.

This observation motivates a two-phase procedure. In the first phase, a standard stochastic modularity optimizer is run repeatedly without constraints, producing a family of trajectories and final partitions distributed within the unconstrained PSM. In the second phase, the information extracted from the best first-phase solutions is converted into a set of \emph{locked groups}, and the optimizer is restarted under these constraints. 
The result is a geometry-aware refinement process: the first phase explores broadly, whereas the second phase searches only within the residual region that remains compatible with the stable structure identified in the first phase.

\paragraph{Why a second phase is useful.} The need for a second phase follows directly from two features of modularity optimisation already visible in the PSM.  First, high-modularity partitions are often concentrated in a limited region of the map, indicating that the full combinatorial space contains a large number of partitions that are not competitive.  Second, stochastic heuristics may still terminate at different local optima inside that region, even when they agree on most of the underlying community structure. 
In such cases, much of the search effort is effectively spent re-discovering the same robust core, while the real uncertainty is confined to a small subset of nodes.  A second phase that freezes the stable part of the partition and optimizes only the uncertain remainder is therefore both natural and well aligned with the geometry of the PSM.

\begin{figure}[t]
    \centering
    \includegraphics[width=\textwidth]{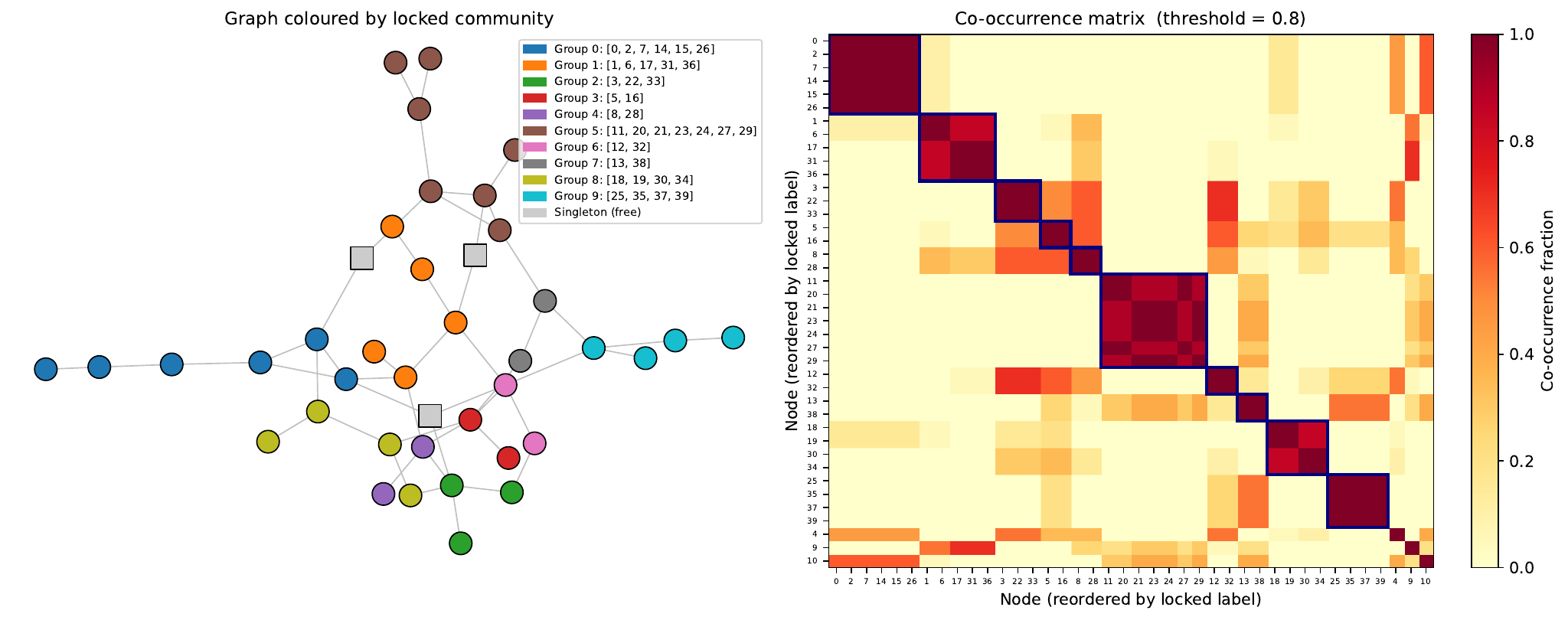}
    \caption{
    Construction of locked groups from the phase~1 ensemble. 
    Left: the network coloured according to the groups obtained by thresholding the co-occurrence matrix derived from the best phase~1 partitions. 
    Nodes belonging to the same locked group share the same colour, while square markers denote nodes that remain unconstrained singletons and may change assignment during the refinement phase. 
    Right: the node co-occurrence matrix $C_{ij}$ computed from the selected ensemble of high-quality partitions, where each entry represents the fraction of runs in which nodes $i$ and $j$ are assigned to the same community. 
    After reordering nodes by locked label, the matrix reveals a block structure corresponding to strongly co-assigned node sets. 
    Applying a threshold (here $\tau=0.8$) identifies the stable groups used to define the locked structure in phase~2.
    }
    \label{fig:cooccurrence_locked_groups}
\end{figure}

Figure~\ref{fig:cooccurrence_locked_groups} illustrates this step for the example network considered in this work.  The co-occurrence matrix obtained from the best phase~1 runs exhibits a clear block structure, indicating groups of nodes that are almost always assigned to the same community. 
Applying a threshold to this matrix produces a set of locked groups that form the backbone of the phase~2 search.
\paragraph{How the constrained region is constructed. } Let $\{\Pi^{(r)}\}_{r=1}^{R}$ denote the final partitions obtained from $R$ independent runs of a stochastic optimizer in phase~1. 
From a selected subset of high-quality runs, we define the co-occurrence matrix
\begin{equation}
    C_{ij} = \frac{1}{R^\ast}\sum_{r=1}^{R^\ast} \mathbf{1}\!\left[c_i^{(r)} = c_j^{(r)}\right],
\end{equation}
where $R^\ast \le R$ is the number of top runs retained and $\mathbf{1}[\cdot]$ is the indicator function. 
Node pairs with $C_{ij}\ge \tau$, for a prescribed threshold $\tau$, are interpreted as consistently co-assigned.  The connected components induced by these high-confidence links define the locked groups used in phase~2, while nodes not absorbed into any non-trivial component remain free singletons.

In the full-matrix convention adopted here, granularity satisfies $\Gamma \in [n,n^2]$, with $\Gamma=n$ for the all-singleton partition.  Once locked groups are introduced, the search can no longer start from $\Gamma=n$, because some pairs are already forced to lie in the same community.  If the locked groups are denoted by $L_1,\dots,L_M$ and the remaining free nodes are singletons, the constrained search begins at
\begin{equation}
    \Gamma_0^{\mathrm{locked}}
    = n + \sum_{a=1}^{M} |L_a|(|L_a|-1)
    = \sum_{a=1}^{M} |L_a|^2 + n_{\mathrm{free}},
\end{equation}
where $n_{\mathrm{free}}$ is the number of unconstrained singleton nodes.  Geometrically, this quantity defines the left boundary of the constrained search region: all partitions explored in phase~2 must satisfy $\Gamma \ge \Gamma_0^{\mathrm{locked}}$.

\begin{figure}[h]
    \centering
    \includegraphics[width=\textwidth]{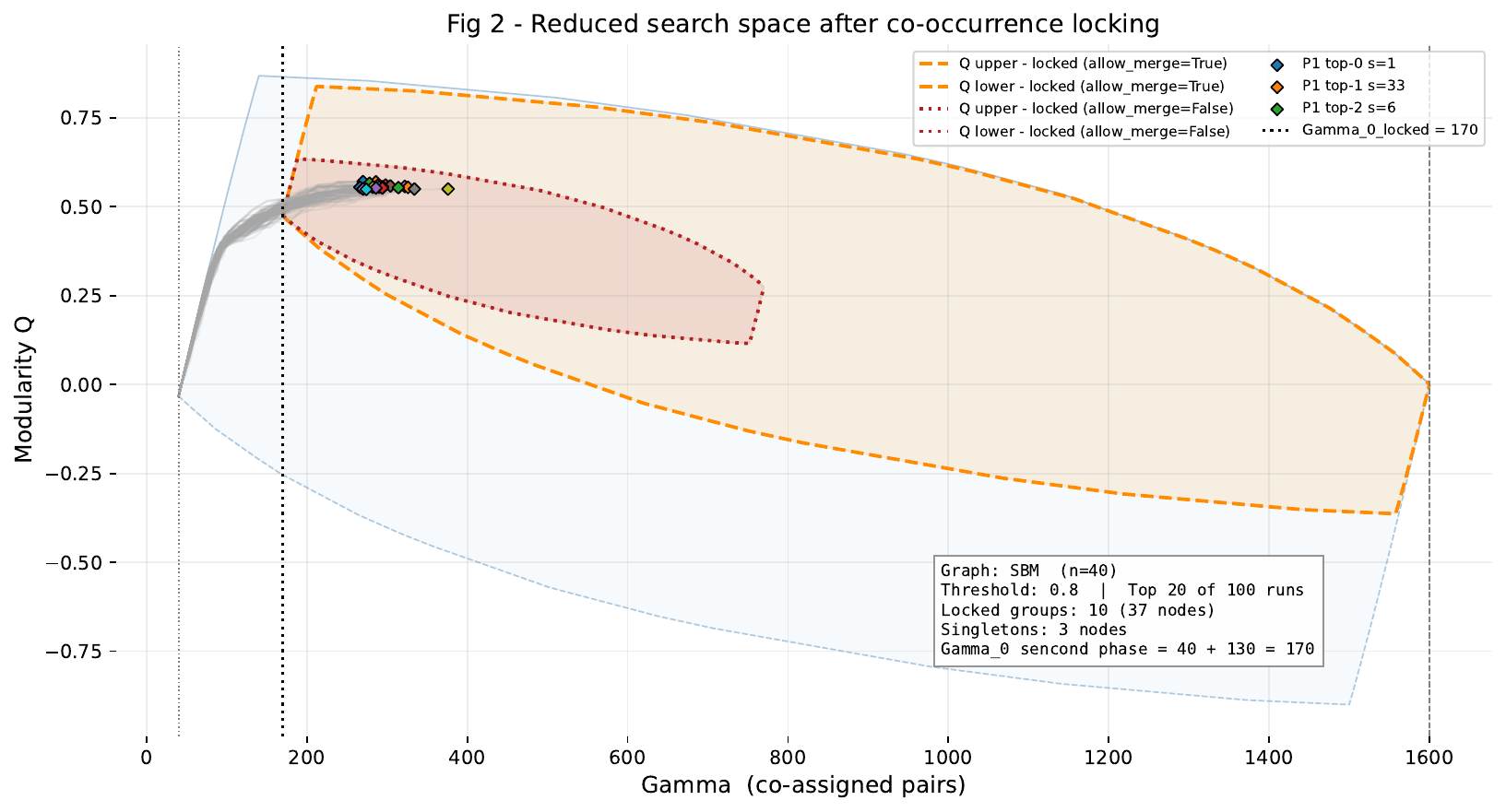}
    \caption{%
    Reduced search space after co-occurrence locking. 
    The grey background corresponds to the original feasible region of the PSM. 
    The vertical dotted line marks the new constrained starting point $\Gamma_0^{\mathrm{locked}}$, induced by the stable co-assignment relations extracted from the best phase~1 solutions. 
    The \emph{outer} constrained envelope (orange) contains all partitions compatible with the locked groups when further mergers are still permitted, whereas the \emph{inner} constrained envelope (red) represents the stricter refinement regime used in phase~2, where locked groups are preserved and only the remaining free singleton nodes are repositioned. 
    The two envelopes therefore distinguish between the residual space implied by the consensus structure and the smaller region actually searched during refinement.
    }
    \label{fig:psm_locked_search_space}
\end{figure}

\paragraph{Double envelopes and residual search space.} The effect of locking can be understood directly in the PSM through a second pair of modularity envelopes. Figure~\ref{fig:psm_locked_search_space} shows two nested constrained regions.  The outer locked envelope corresponds to the set of partitions that respect the locked intra-group relations while still allowing those locked groups to merge with one another. 
This defines the broadest search space consistent with the phase~1 consensus.  The inner locked envelope is more restrictive: it corresponds to the refinement step actually used here, in which the locked groups are preserved, and only the free singleton nodes are allowed to change assignment.  Hence, the two envelopes represent two distinct levels of conditioning induced by the same empirical consensus structure.

This distinction is important conceptually.  The outer envelope answers the question: \emph{what part of partition space remains feasible once the robust co-assignment relations discovered in phase~1 are enforced?} 
The inner envelope answers the algorithmic question: \emph{what part of that residual space is actually explored by the refinement procedure adopted in phase~2?} 
In this sense, the constrained PSM does not merely visualize the output of the second phase; it explicitly defines the admissible region in which that search takes place.

\begin{figure}[h]
    \centering
    \includegraphics[width=\textwidth]{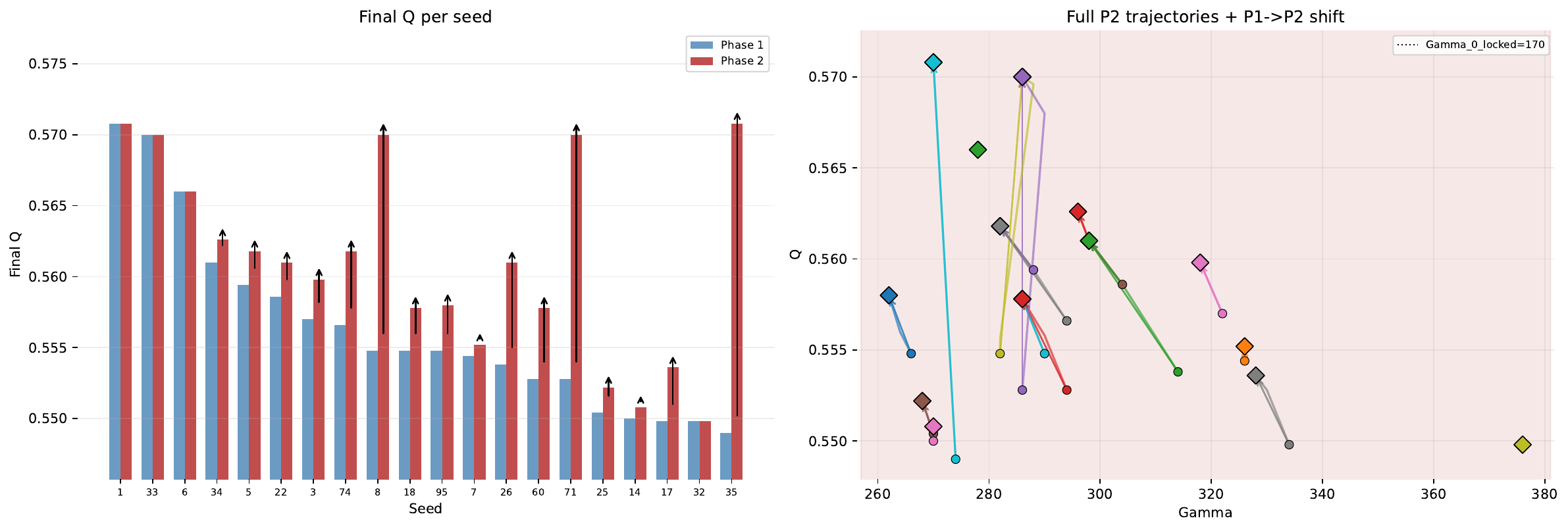}
    \caption{%
    Comparison between the unconstrained and constrained phases for the top seeds retained from phase~1. 
    Left: final modularity values before and after refinement; arrows highlight cases in which phase~2 improves upon the corresponding phase~1 solution. 
    Right: trajectories within the constrained PSM region, together with the phase~1 $\rightarrow$ phase~2 displacement. 
    The figure shows that refinement can improve modularity even for solutions that appear relatively weak at the end of the first phase, indicating that much of the residual error may be concentrated in a small number of ambiguous nodes rather than in the global community backbone.
    }
    \label{fig:psm_phase_comparison}
\end{figure}

\paragraph{Example: modularity improvement in the second phase.} In the example considered here, the second phase produces a clear improvement over the first-phase solutions. The constrained search does not simply reproduce the same partitions under a different initialisation; rather, it exploits the reduced search space to refine the placement of the few nodes that remain uncertain after consensus locking.  This is visible in Figure~\ref{fig:psm_phase_comparison}, where the final modularity values obtained after refinement are compared to the corresponding phase~1 outcomes for the same seeds.

A particularly instructive aspect of the figure is that seeds that appear relatively weak after the unconstrained search may still possess substantial room for improvement once the robust core structure has been fixed.  In other words, a modest phase~1 outcome does not necessarily indicate that the run is globally uninformative: it may already contain the correct large-scale organisation, while failing only in the treatment of a small number of ambiguous nodes. 
The second phase isolates this residual uncertainty.  Accordingly, one of the weaker first-phase outcomes in Figure~\ref{fig:psm_phase_comparison} undergoes one of the largest upward corrections after refinement, showing that the difficult part of the problem can be highly localized even when the initial modularity is not among the best.

The same figure also shows that the phase~1 to phase~2 displacement is typically small in $\Gamma$ relative to the full width of the original PSM, but non-negligible in $Q$.  This is precisely the behavior expected from a successful refinement method: most of the partition structure is retained, while the remaining degrees of freedom are used to move the solution upward within the constrained feasible region.  For the present example, the best modularity achieved in phase~2 exceeds the best value found in phase~1, demonstrating that the PSM-inspired restriction is not only descriptive but also operationally useful for optimisation.

\paragraph{Interpretation.}
This two-phase scheme should not be viewed as a replacement for existing modularity heuristics, but rather as a PSM-guided wrapper around them. 
Its purpose is to convert the geometric information revealed by repeated exploratory runs into a principled restriction of partition space. 
In this sense, the method extends the role of the PSM from a diagnostic representation to an algorithmic guide: phase~1 estimates where the relevant region of partition space lies, and phase~2 concentrates the search within that region. 
The approach is especially appealing for networks in which the optimizer repeatedly identifies a stable community backbone together with a small set of unresolved or weakly attached nodes, since in that regime the PSM makes the separation between robust structure and residual ambiguity geometrically explicit.

\section{Discussion}
\label{sec:discussion}

The primary contribution of this work is a set of complementary tools that provide visual and analytical intuition for exploring the partition space of complex networks.

The Partition Space Map projects algorithmic outputs onto the $(\Gamma, Q)$ plane, offering a geometric view of how different algorithms, parameter choices, and stochastic initialisations distribute solutions across the feasible region. Unlike scalar comparisons of modularity scores, the PSM reveals the \emph{direction} in which solutions differ---toward finer or coarser granularity, higher or lower quality---making algorithmic behaviour more interpretable. When an algorithm is run repeatedly under different random seeds, the resulting cluster or scatter in the $(\Gamma, Q)$ plane directly exposes its sensitivity to initialisation, as illustrated by the contrasting footprints of Louvain and Label Propagation in Fig.~7.

The modularity bounding curves, derived solely from the sorted entries of the deviation matrix, frame this landscape at minimal computational cost. For any network---including those with no known ground truth---the envelopes provide the shape and size of the feasible region, establishing an absolute reference against which any candidate partition can be positioned.

The PSM has an inherent limitation: partitions that differ structurally but share the same $(\Gamma, Q)$ coordinates collapse to a single point. Modularity plateaus are thus invisible in the map. The Restricted Growth Sequence addresses precisely this gap. By assigning each partition a unique canonical label, the RGS enables exact comparison without pairwise similarity measures such as NMI, and allows researchers to enumerate and explore the partitions that populate a modularity plateau.

The three tools are complementary: the bounding curves frame the feasible region, the PSM reveals how solutions are distributed within it, and the RGS unpacks the degeneracy that the PSM compresses. Together, they transform the abstract, high-dimensional partition space into a structured representation that supports algorithm diagnostics, parameter selection, and characterisation of solution spaces in community detection~\cite{Morea2025}.

Future work may extend the framework in several directions. A natural next step is validation on real-world networks, where ground truth is absent and the bounding curves become the primary reference for understanding the partition space. For algorithm developers, the PSM could be used to trace the full optimisation path, from random initialisation through intermediate steps to final solution, potentially revealing how different heuristics behave, where they stall or diverge. Further extensions may include the exploration of alternative quality functions beyond $Q$. The partition space $\mathcal{P}$ does not carry any natural metric, and modularity is only one of many possible projections; other quality functions or distance measures may reveal complementary aspects of its structure that the $(\Gamma, Q)$ plane leaves hidden.

\section*{Data availability}
All data used in this study and code for computing the PSM is available in GitHub: \href{ https://github.com/fabio-morea/partition-space}{ github.com/fabio-morea/partition-space}

\printbibliography

\end{document}